\documentclass{m2an}

\usepackage{graphicx} 
\usepackage{amsmath,amsthm,amssymb,bm,xcolor,mathtools}
\usepackage[colorlinks=true,citecolor=blue]{hyperref}
\numberwithin{equation}{section}

\begin{document}
\title[Noisy nonlocal aggregation model]{Noisy nonlocal aggregation model with gradient flow structures}
\author{Su Yang}\address{Department of Mathematics and Statistics, University of Massachusetts Amherst, Amherst, MA 01003-4515, USA}
\author{Weiqi Chu}\sameaddress{1}
\author{Panayotis G. Kevrekidis}
\sameaddress{1}
\address{Department of Physics, University of Massachusetts Amherst, Amherst, MA 01003-4515, USA}

%
\begin{abstract}
Interacting particle systems provide a fundamental framework for modeling collective behavior in biological, social, and physical systems. In many applications, stochastic perturbations are essential for capturing environmental variability and individual uncertainty, yet their impact on long-term dynamics and equilibrium structure remains incompletely understood, particularly in the presence of nonlocal interactions. 
We investigate a stochastic interacting particle system governed by potential-driven interactions and its continuum density formulation in the large-population limit.
We introduce an energy functional and show that the macroscopic density evolution has a gradient-flow structure in the Wasserstein-2 space.
The associated variational framework yields equilibrium states through constrained energy minimization and illustrates how noise regulates the density and mitigates singular concentration.
We demonstrate the connection between microscopic and macroscopic descriptions through numerical examples in one and two dimensions.
Within the variational framework, we compute energy minimizers and perform a linear stability analysis.
The numerical results show that the stable minimizers agree with the long-time dynamics of the macroscopic density model.
\end{abstract}

\maketitle

\section{Introduction}

Systems of interacting particles model the dynamics of a collection of individuals that evolve through pairwise interactions and external fields. 
Such models arise in a broad range of scientific contexts, including cell migration and aggregation in biological systems, opinion dynamics in the social sciences, as well as applications in ecology and statistical physics~\cite{castellano2009statistical,motsch2014heterophilious,kolokolnikov2011stability}; see also the books recapping the progress in this field~\cite{NaldiPareschiToscani2010,BellomoDegondTadmor2017,BellomoDegondTadmor2019}.
Within this framework, interaction kernels encode attraction and repulsion mechanisms that regulate interactions across spatial or opinion distances.
At the population level, these models give rise to a variety of emergent behaviors, including clustering, pattern formation, spreading, as well as bifurcation phenomena between different patterns and transitions between ordered and disordered states~\cite{lorenz2005stabilization,bernoff2011primer,mogilner2003mutual}.

Stochastic perturbations arise naturally in interacting particle systems as a means to represent environmental fluctuations, random decision-making, and unresolved microscale effects.
From a modeling perspective, noise introduces an additional mechanism that competes with pairwise interaction forces and can substantially influence collective behavior.
For instance, in opinion dynamics models with bounded-confidence interactions \cite{hegselmann2015opinion}, deterministic dynamics drive the density toward singular measures, in the form of sums of Dirac delta functions, whereas stochasticity smooths sharp clusters and alters the structure and stability of long-term dynamics \cite{wang2015noisyhegselmannkrausesystemsphase,pineda2013noisy}.
By tuning the noise level in a manner analogous to the role of temperature in thermodynamical systems, the system can vary between disordered dynamics and convergence toward fixed points, identifying phase transitions along the way \cite{wang2015noisyhegselmannkrausesystemsphase,castellano2009statistical}.
More broadly, noisy extensions of interacting-agent models continue to attract significant interest, as they capture more realistic dynamics and reveal noise-induced transitions that do not arise in purely deterministic settings \cite{helfmann2023modelling,wehlitz2025approximating,chen2019heterogeneous}.
The presence of noise and random initial conditions increases variability at the agent level, which complicates the extraction of macroscopic statistical behavior from individual trajectories and motivates the use of continuum descriptions and ensemble-based analysis~\cite{vicsek,garnier2016consensusconvergencestochasticeffects}.

Microscopic interacting particle systems and their stochastic extensions have a complementary macroscopic description in the limit of infinitely many particles.
Under suitable assumptions on the interaction kernel and scaling parameters, results from mean-field theory and propagation of chaos show that the empirical distribution of agent states converges to the law of a representative mean-field particle \cite{sznitman2006topics,chaintron2022propagation,dawson1983critical,garnier2016consensusconvergencestochasticeffects}.
The limiting dynamics can be described either by the evolution of this mean-field particle coupled to its own distribution or by kinetic equations (e.g., Fokker--Planck equations) characterized by the probability density function of the mean-field particle.
The macroscopic formulation provides direct access to population-level dynamics and avoids the need for repeated simulations to obtain statistical ensembles of systems with a large number of interacting particles~\cite{sznitman2006topics}.
In the presence of stochastic noise and random initial conditions, this macroscopic viewpoint becomes particularly useful, since extracting collective behavior from individual trajectories entails substantial computational cost and typically requires averaging over many realizations to reduce sampling variability \cite{del2013mean}. 

Interacting particle systems with pairwise interactions and stochastic perturbations exhibit a wide range of collective behaviors. 
The dynamics depend on the form of the interaction kernel, the relative strength between the potentially featured attraction and repulsion therein, and the noise level, leading to density outcomes such as spreading to infinity, the formation of localized profiles, concentration into (one or more) discrete clusters, or transitions between distinct configurations \cite{leverentz2009asymptotic}.
When interaction forces arise from some underlying potential functions, the population density evolves as a gradient flow, driven by the dissipation of an associated energy.
Therefore, the equilibrium states correspond to stationary configurations that minimize the energy under mass and nonnegativity constraints, and this energy framework provides a systematic approach to understanding long-term dynamics in both noisy and noiseless interacting particle systems \cite{bernoff2011primer,canizo2015existence}. 
In the absence of noise, several studies develop variational frameworks for continuum nonlocal interaction models and derive equilibrium states by minimizing the energy \cite{bernoff2011primer,canizo2015existence}.
These works show that equilibria of noiseless systems are compactly supported and may feature singular structures, such as Dirac delta concentrations or jump discontinuities at the boundary of the support \cite{fetecau2011swarm,bernoff2011primer}.
In contrast, noise alters the singular densities by regularizing equilibria and suppressing singular concentration.

In this work, we study noisy nonlocal aggregation models and extend variational and dynamical analysis to noisy nonlocal aggregation models. We describe the models at both the microscopic and macroscopic levels, starting from a system of interacting particles subject to stochastic perturbations and proceeding to a continuum formulation for the population density in the large-population limit.
We introduce an energy functional that combines nonlocal interactions, external forces, and an entropy term induced by noise. We show that the macroscopic density equation has a gradient-flow formulation in the Wasserstein-2 space.
Using a variational framework in the probability space, we derive conditions that characterize energy minimizers and establish their connection to stationary states. We perform a numerical linear stability analysis of these energy minimizers and analyze the long-term behavior of the system in the presence of noise.

We organize the paper as follows.
In Section~\ref{sec: model}, we introduce the noisy nonlocal aggregation model through both microscopic and macroscopic descriptions and discuss their connections.
In Section~\ref{sec: energy}, we show that the macroscopic model has a gradient-flow formulation in the Wasserstein-2 space and derive sufficient conditions to minimize the energy from a variational perspective.
In Section~\ref{sec: numerics}, we describe numerical methods for solving the microscopic model in the form of stochastic differential equations (SDEs) and the macroscopic model in the form of a nonlocal aggregation diffusion equation, and we outline a procedure for comparing density evolution in the large-population limit.
In Section~\ref{sec: examples}, we perform numerical studies of the noisy nonlocal aggregation model with three representative interaction potentials in one and two dimensions.
We conclude and offer a number of directions for future studies in Section~\ref{sec: summary}.

\section{A nonlocal aggregation model with noise}
\label{sec: model}
\subsection{Model description from microscopic and macroscopic perspectives}
We consider a system of $N$ interacting agents and represent the state of the $i$th agent at time $t$ by a vector $\bm{x}_i(t)\in\Omega\subset\mathbb{R}^d$.  
Each agent adjusts its state through interaction forces $\bm{\theta}$, an external force $\bm{f}$, and stochastic fluctuations modeled by Brownian motion.  
The agent-based dynamics follow the SDEs
\begin{equation} \label{eq: SDE_model}
d\bm{x}_i(t)
= -\frac{1}{N}\sum_{j=1}^N \bm{\theta}(\bm{x}_i(t)-\bm{x}_j(t))\,dt
  - \bm{f}(\bm{x}_i(t))\,dt
  + \sigma\,d\bm{W}_t^i,
\qquad i=1,\dots,N,
\end{equation}
where $\sigma>0$ is the diffusion coefficient and $\{\bm{W}_t^i\}_{i=1}^N$ are independent $d$-dimensional Wiener processes.

To study collective behaviors of the whole population, we introduce the empirical measure
\begin{equation}\label{eq: rhoN}
\rho^N(\bm{x},t)
= \frac{1}{N}\sum_{j=1}^N \delta_{\bm{x}_j(t)}(\bm{x}),
\end{equation}
and rewrite \eqref{eq: SDE_model} using the empirical measure notation as
\begin{equation}
d\bm{x}_i(t)
= \bm{b}(\bm{x}_i(t),\rho^N(\bm{x},t))\,dt
  + \sigma\,d\bm{W}_t^i,
\end{equation}
where $\bm{b}(\bm{x},\rho)$ is the drift term, defined as
\begin{equation} \label{eq: b}
\bm{b}(\bm{x},\rho)
= -(\bm{\theta}\star\rho)(\bm{x}) - \bm{f}(\bm{x}),
\end{equation}
and $(\bm{\theta}\star\rho)(\bm{x}) = \int_{\Omega} \bm{\theta}(\bm{x}-\bm{y})\,\rho(\bm{y})\,d\bm{y}$ is the convolution, defined element-wisely.

When the initial states $\bm{x}_1(0),\dots,\bm{x}_N(0)$ are i.i.d. with the same law, the theory of propagation of chaos~\cite{chaintron2022propagation} shows that, as $N\rightarrow\infty$, any finite collection of particles become asymptotically independent and each particle becomes asymptotically distributed according to the same one-particle law. In this limit, a representative particle satisfies the nonlinear McKean--Vlasov SDE
\begin{equation} \label{eq: limiting}
d\bm{x}(t)
= \bm{b}(\bm{x}(t),\rho(\bm{x},t))\,dt + \sigma\,d\bm{W}_t,
\end{equation}
where $\rho(\bm{x},t)$ is the law of $\bm{x}(t)$.
The density $\rho(\bm{x},t)$ follows the Fokker--Planck equation ~\cite{gartner1988mckean,garnier2016consensusconvergencestochasticeffects} associated with \eqref{eq: limiting}, which is
\begin{equation}
\partial_t \rho(\bm{x},t)
= -\nabla\cdot\!\left(\rho(\bm{x},t)\,\bm{b}(\bm{x},\rho(\bm{x},t))\right)
  + \frac{\sigma^2}{2}\Delta\rho(\bm{x},t),
\end{equation}
where $\Delta=\sum_{j=1}^{d}\partial^2_{x_j}$.  
We substitute \eqref{eq: b} into the above equation and obtain the nonlocal aggregation diffusion equation
\begin{equation} \label{eq: pde_model}
\partial_t \rho
= \nabla\cdot\!\left(\rho \, (\bm{\theta}\star\rho + \bm{f})\right)
  + \frac{\sigma^2}{2}\Delta\rho,
\end{equation}
which is the PDE model describing the interacting particle system in the mean-field limit \cite{garnier2016consensusconvergencestochasticeffects}.

\subsection{Energy dissipation in the nonlocal aggregation model} 

Stochastic noise and random initial states make it difficult to extract collective statistical behavior directly from agent-based simulations, especially for systems with a large number of particles.
The connection between the microscopic description \eqref{eq: SDE_model} and the macroscopic description \eqref{eq: pde_model} provides a natural framework for studying collective behavior in the large-population regime.  
In the rest of the paper, we focus on the macroscopic, density-based description, with particular attention to its energy structure when the forces are in gradient forms.

Suppose that the interaction force $\bm{\theta}$ and the external force $\bm{f}$ satisfy
\begin{equation}\label{eq: fassumption}
\bm{\theta} = \nabla \Theta, 
\quad 
\bm{f} = \nabla F,
\end{equation}
where $\Theta$ and $F$ are scalar potentials. 
Let $\rho$ be a probability density on $\Omega$ that satisfies mass conservation and nonnegativity,
\begin{equation}
\int_{\Omega} \rho(\bm{x})\,d\bm{x} = 1,
\quad
\rho(\bm{x}) \ge 0 \ \ \text{for all } \bm{x}\in\Omega .
\end{equation}
We define the energy functional $\mathcal{E}[\rho]$ by
\begin{equation}\label{eq: energy}
\mathcal{E}[\rho]
=
\frac{1}{2}\int_{\Omega^2}
\rho(\bm{x})\,\rho(\bm{y})\,\Theta(\bm{x}-\bm{y})\,d\bm{x}\,d\bm{y}
+ \int_{\Omega} \rho(\bm{x})\,F(\bm{x})\,d\bm{x}
+ \frac{\sigma^2}{2}\int_{\text{supp}(\rho)}
\rho(\bm{x})\ln\rho(\bm{x})\,d\bm{x},
\end{equation}
where $\text{supp}(\rho)$ is the compact support of $\rho$.
The first term represents the interaction energy arising from continuum limits of symmetric pairwise forces and has been widely used to characterize nonlocal interaction energies in aggregation models \cite{bernoff2011primer,canizo2015existence}, the second term represents the external potential energy, and the last term corresponds to the entropy induced by stochastic forcing~\cite{jordan,ambrosio2005gradient}.

Let $\rho=\rho(\bm{x},t)$ be a solution of the PDE model \eqref{eq: pde_model} and $A=\text{supp}(\rho)$ be the compact support of $\rho$. 
An important property of the system is that the energy functional $\mathcal{E}[\rho]$ decreases in time, as discussed in \cite{carrillo2015finite}.
For completeness, we outline the derivation below.
For notational simplicity, we omit the explicit dependence on $\bm{x}$ in the integrals and compute the time derivative of the energy:
\begin{subequations} \label{eq: energy_decay}
    \begin{align}
        \frac{d}{dt}\mathcal{E}[\rho] 
        &= \int_A\nabla\cdot\left(\rho\left(\bm{\theta}\star\rho\right) + \rho \bm{f}+\frac{\sigma^2}{2}\nabla\rho\right)\left(\Theta\star\rho+F+\frac{\sigma^2}{2}\left(1+\ln\rho\right)\right)\,d\bm{x} \\
        &= - \int_A\left(\rho\left(\bm{\theta}\star\rho\right)+\rho\bm{f}+\frac{\sigma^2}{2}\nabla\rho\right)
        \cdot 
        \nabla\left(\Theta\star\rho+F+\frac{\sigma^2}{2}\ln\rho\right)\,d\bm{x} \label{gradient}\\
        &= -\int_A\rho\left|\bm{\theta}\star\rho+\bm{f}+\frac{\sigma^2}{2}\nabla\ln\rho\right|^2\,d\bm{x} \leq 0.
    \end{align}
\end{subequations}
The integration by parts step in \eqref{gradient} requires a suitable boundary condition that removes boundary contributions in the energy balance.  
For a finite domain $\Omega$, this condition holds under no-flux boundary conditions or periodic boundary conditions when $\Omega$ is the torus $\left[-L,L\right]^d$. 
When $\Omega=\mathbb{R}^d$, this condition holds naturally as the density $\rho$ decays sufficiently fast as $|\bm{x}|$ goes to infinity.  
Under these settings, the boundary term vanishes, and the energy dissipation identity remains valid.

\section{Energy minimization in the probability space}
\label{sec: energy}
\subsection{Gradient-flow formulation in the Wasserstein-2 space}
In addition to the energy dissipation property, the PDE model \eqref{eq: pde_model} corresponds to a gradient flow in the Wasserstein-2 space. 
In the Wasserstein-2 space, the generalized gradient of a functional $E$ takes the form \cite{ambrosio2005gradient}
\begin{equation} \label{eq: w2gradient}
\nabla_{W_2} E[\rho]
=
- \nabla\cdot\!\left(
\rho \nabla \frac{\delta E}{\delta\rho}
\right),
\end{equation}
where $\delta E/\delta\rho$ is the first variation density of $E$~\cite{RevModPhys.70.467}.

Let $\mathcal{E}$ be the energy defined in \eqref{eq: energy}. 
Direct computation yields
\begin{equation} \label{eq: deltaE}
\frac{\delta \mathcal{E}}{\delta\rho}
=
\Theta\star\rho + F
- \frac{\sigma^2}{2}\,\frac{\delta \mathcal{S}}{\delta\rho},
\end{equation}
where $\mathcal{S}$ is the entropy functional
\begin{equation}\label{eq: entropy functional}
\mathcal{S}[\rho]
=
-\int_{\text{supp}(\rho)} \rho(\bm{x})\ln\rho(\bm{x})\,d\bm{x}.
\end{equation}
Let $\tilde{\rho}$ be an admissible perturbation such that
\begin{equation} \label{eq: nonnegative}
\rho_\varepsilon(\bm{x}) = \rho(\bm{x}) + \varepsilon \tilde{\rho}(\bm{x}) \ge 0
\quad \text{for all } \bm{x} \in \Omega
\end{equation}
for sufficiently small $|\varepsilon|$. 
Since $\varepsilon$ can be either positive or negative, the nonnegativity requirement \eqref{eq: nonnegative} implies that $\tilde{\rho}(\bm{x}) = 0$ whenever $\rho(\bm{x}) = 0$. 
Hence, we have $\operatorname{supp}({\rho}_\varepsilon)=\operatorname{supp}(\rho)$, which implies the integration domain remains the same for a fixed $\rho$ (independent of $\varepsilon$). 
We compute the functional derivative of $\mathcal{S}$ and obtain ${\delta\mathcal{S}}/{\delta\rho}=-(\ln\rho+1)$~\cite{RevModPhys.70.467}.
Using the gradient definition \eqref{eq: w2gradient} and the variational form \eqref{eq: deltaE}, we have 
\begin{equation}
\begin{aligned}
\nabla_{W_2} \mathcal{E}[\rho]
&=
-\nabla\cdot\!\left(
\rho\nabla\left(
\Theta\star\rho + F + \frac{\sigma^2}{2}(\ln\rho+1)
\right)
\right)=
-\nabla\cdot\!\left(\rho(\bm{\theta}\star\rho+\bm{f})\right)
- \frac{\sigma^2}{2}\Delta\rho.
\end{aligned}
\end{equation}
Therefore, the PDE model \eqref{eq: pde_model} can also be written as
\begin{equation}
\partial_t \rho = - \nabla_{W_2} \mathcal{E}[\rho],
\end{equation}
which identifies \eqref{eq: pde_model} as a gradient flow with respect to the energy $\mathcal{E}$ in the Wasserstein-2 space.

\subsection{Energy minimization in the probability space} 
Since the PDE model \eqref{eq: pde_model} has a gradient-flow structure, its stationary solutions correspond to probability densities that minimize the energy $\mathcal{E}$ \eqref{eq: energy} over the space of admissible probability measures. 
In this subsection, we derive sufficient conditions to minimize the energy $\mathcal{E}$ under mass and nonnegativity constraints and analyze how the potential and entropy terms influence the energy minimizers in the probability space.  
The energy minimization problem for the nonlocal aggregation model without noise is studied in \cite{bernoff2011primer}.

Let $\rho_{\varepsilon} = \rho + \varepsilon \tilde{\rho}$ be a perturbed density with $\varepsilon>0$, satisfying the mass conservation and nonnegative properties.
Expanding the energy $\mathcal{E}$ around $\varepsilon=0^+$ yields
\begin{equation}\label{eq: p}
\mathcal{E}[\rho+\varepsilon\tilde{\rho}]
=
\mathcal{E}[\rho]
+ \varepsilon\, \mathcal{E}_1[\rho,\tilde{\rho}]
+ \varepsilon^2\, \mathcal{E}_2[\rho,\tilde{\rho}]
+ o(\varepsilon^2).
\end{equation}
Let $\rho^*$ be a local minimizer of the energy $\mathcal{E}[\rho]$.  
To find the governing equations that $\rho^*$ satisfies, we consider two types of admissible perturbations $\tilde{\rho}$.
First, consider perturbations $\tilde{\rho}$ whose support lies within the support of $\rho^*$, that is, $\text{supp}(\tilde{\rho}) \subseteq \text{supp}(\rho^*)$.
Under this condition, both $\tilde{\rho}$ and $-\tilde{\rho}$ are admissible perturbations.  
The local minimizer $\rho^*$ must satisfy
\begin{equation}\label{eq: constraint1}
\mathcal{E}_1(\rho^*,\tilde{\rho}) = 0,
\quad
\mathcal{E}_2(\rho^*,\tilde{\rho}) \ge 0,
\qquad
\text{for all admissible } \tilde{\rho}
\text{ with supp}(\tilde{\rho})\subseteq \text{supp}(\rho^*).
\end{equation}
Next, consider perturbations $\tilde{\rho}$ whose support is not contained in $\text{supp}(\rho^*)$.  
In this case, admissibility requires $\tilde{\rho}(\bm{x}) > 0$ for $\bm{x}\in\text{supp}(\tilde{\rho}) \setminus \text{supp}(\rho^*)$.  
For such perturbations, the first variation must satisfy
\begin{equation} \label{eq: constraint2}
\mathcal{E}_1(\rho^*,\tilde{\rho}) > 0, \qquad 
\text{for all admissible } \tilde{\rho}
\text{ with }\tilde{\rho}(\bm{x})>0 \text{ on } \text{supp}(\tilde{\rho}) \setminus \text{supp}(\rho^*).
\end{equation}

For the first type of perturbations in \eqref{eq: constraint1}, the first order variation takes the form
\begin{equation} \label{eq: first}
\mathcal{E}_1[\rho,\tilde{\rho}]
=
\int_{\text{supp}(\rho^*)}
\tilde{\rho}(\bm{x})
\left(
(\rho\star\Theta)(\bm{x})
+ F(\bm{x})
+ \frac{\sigma^2}{2}\big(\ln\rho(\bm{x})+1\big)
\right)
\,d\bm{x}.
\end{equation}
Since $\tilde{\rho}$ is arbitrary and of zero mass, the first-order optimality condition implies that there exists a scalar $\lambda\in\mathbb{R}$ such that 
\begin{equation}\label{eq: condition1a}
(\rho^*\star\Theta)(\bm{x})
+ F(\bm{x})
+ \frac{\sigma^2}{2}\ln\rho^*(\bm{x})
= \lambda
\end{equation}
for $\bm{x}\in\text{supp}(\rho^*)$~\cite{bernoff2011primer}.
The second-order variation in \eqref{eq: constraint1} has the form
\begin{equation} \label{eq: E2}
    \mathcal{E}_2[\rho,\tilde{\rho}]
=
\frac{1}{2}
\int_{\Omega^2}
\tilde{\rho}(\bm{x})\tilde{\rho}(\bm{y})
\Theta(\bm{x}-\bm{y})
\,d\bm{x}\,d\bm{y}
+
\frac{\sigma^2}{4}
\int_{\text{supp}(\rho)}
\frac{\tilde{\rho}^2(\bm{x})}{\rho(\bm{x})}
\,d\bm{x}.
\end{equation}
Assume that $\rho^* \in L^{\infty}(\Omega)$ with the infinite norm $\|\cdot\|_{\infty}$. 
Since $\text{supp}(\tilde{\rho})\subseteq\text{supp}(\rho)$, we have
\begin{equation}\label{eq: explanation of condition 2}
\begin{aligned}
\mathcal{E}_2[\rho^*,\tilde{\rho}]
&\ge
\frac{1}{2}
\int_{\Omega}
\tilde{\rho}(\bm{x})\big(\tilde{\rho}\star\Theta\big)(\bm{x})
\,d\bm{x}
+
\frac{\sigma^2}{4\|\rho^*\|_{\infty}}
\int_{\Omega}
\tilde{\rho}^2(\bm{x})
\,d\bm{x}.
\end{aligned}
\end{equation}
We apply Parseval's identity and obtain
\begin{equation}
\begin{aligned}
\mathcal{E}_2[\rho^*,\tilde{\rho}]
\ge \frac{1}{2}
\int_{\mathbb{R}^d}
\big|\hat{\tilde{\rho}}(\bm{k})\big|^2
\left(
\hat{\Theta}(\bm{k})
+
\frac{\sigma^2}{2\|\rho^*\|_{\infty}}
\right)
\,d\bm{k},
\end{aligned}
\end{equation}
where $\hat{\cdot}$ denotes the Fourier transform and $\bm{k}$ is the Fourier wavevector.
Consequently, one sufficient condition to guarantee the second-order condition in \eqref{eq: constraint1} is the spectral positivity condition
\begin{equation}\label{eq: condition1b}
\hat{\Theta}(\bm{k})
+
\frac{\sigma^2}{2\|\rho^*\|_{\infty}}
\ge 0
\quad
\text{for all } \bm{k}\in\mathbb{R}^d.
\end{equation}
Together, \eqref{eq: condition1a} and \eqref{eq: condition1b} form the constraints of the minimizer $\rho^*$ with respect to the first type of perturbations.

Now, we turn to the constraint in \eqref{eq: constraint2} for the second type of perturbations. 
We claim that in the presence of noise (i.e., $\sigma>0$), the minimizer $\rho^*$ must satisfy that $\text{supp}\left(\rho^*\right) = \Omega$, resulting in the second type of perturbations not existing. 
To show this, we examine the variational form of the entropy functional $\mathcal{S}[\rho]$ \eqref{eq: entropy functional}. 
Let $A=\text{supp}(\rho^*)$.  
We examine the one-sided directional derivative of the entropy. Direct computations yield
\begin{equation}\label{eq: entropy argument}
\begin{aligned}
&\lim_{\varepsilon\to 0^+}
\frac{\mathcal{S}[\rho^*+\varepsilon\tilde{\rho}]-\mathcal{S}[\rho^*]}{\varepsilon} \\
=&
-\lim_{\varepsilon\to 0^+}
\int_{A}
\frac{(\rho^*+\varepsilon\tilde{\rho})\ln(\rho^*+\varepsilon\tilde{\rho})-\rho^*\ln\rho^*}{\varepsilon}
\,d\bm{x}
-
\lim_{\varepsilon\to 0^+}
\int_{\Omega\setminus A}
\tilde{\rho}(\bm{x})\ln(\varepsilon\tilde{\rho}(\bm{x}))
\,d\bm{x} \\ 
=&
-
\int_{A}
\tilde{\rho}(\bm{x})\big(\ln\rho^*(\bm{x})+1\big)
\,d\bm{x}
-
\int_{\Omega\setminus A}
\tilde{\rho}(\bm{x})\ln\tilde{\rho}(\bm{x})
\,d\bm{x}
-
\lim_{\varepsilon\to 0^+}\ln(\varepsilon)
\int_{\Omega\setminus A}
\tilde{\rho}(\bm{x})\,d\bm{x}.
\end{aligned}
\end{equation}
Notice that $\int_{\Omega\setminus A}\tilde{\rho}(\bm{x})\,d\bm{x} \ge 0$ for the second type of perturbations $\tilde{\rho}$. Since $\lim_{\varepsilon\to 0^+}\ln(\varepsilon)=-\infty$, the limit in \eqref{eq: entropy argument} diverges to $\infty$ when $\int_{\Omega\setminus A}\tilde{\rho}(\bm{x})\,d\bm{x} > 0$, leading the first order variation in \eqref{eq: constraint2} to $-\infty$. 
Therefore, the variational inequality in \eqref{eq: constraint2} holds only if $\int_{\Omega\setminus A}\tilde{\rho}(\bm{x})\,d\bm{x} = 0$.
Because admissible perturbations satisfy $\tilde{\rho}(\bm{x})>0$ on $\Omega\setminus A$, this condition forces $\Omega\setminus A$ to be an empty set, resulting in the second type of perturbations not existing when $\sigma>0$.

\section{Numerical studies for the noisy nonlocal aggregation models}
\label{sec: numerics}
In this section, we illustrate the proposed framework using one- and two-dimensional examples and numerically study the corresponding dynamics and associated energy minimizers.
We introduce numerical methods for solving the SDE model \eqref{eq: SDE_model} and the PDE model \eqref{eq: pde_model} and compare the corresponding densities in the large population limit, where we expect a good agreement between the two, as discussed above.
We also develop numerical approaches to compute the associated energy minimizers in the probability space.  
In addition, we briefly discuss the results of linear stability analysis around the computed energy minimizers, corroborating the spectral stability properties of the stationary states we obtain.  

\subsection{Numerical methods for solving the SDE and PDE models}

We apply the Euler--Maruyama method to solve the SDE model \eqref{eq: SDE_model} for its simplicity, but one can also use stochastic Runge--Kutta methods to achieve higher order accuracy.
We generate the initial states $\{\bm{x}_i(0)\}_{i=1}^N$ independently from a prescribed initial density $\rho_0(\bm{x})$ and evolve the system forward in time using a fixed time step.  
To reduce statistical fluctuations induced by stochastic forcing, we repeat the simulation multiple times and collect an ensemble of independent realizations of the agent dynamics. 
At each time $t$, we approximate the macroscopic density by constructing a discrete empirical density $\rho^d$ with a similar idea as the empirical measure $\rho^N$ in \eqref{eq: rhoN}.  
In particular, we partition the spatial domain $\Omega$ into a finite collection of disjoint bins $\{B_k\}_{k=1}^K$ with volume $|B_k|$.  
We then approximate the density by
\begin{equation}\label{eq: rhod}
\rho^d(\bm{x},t)
= \sum_{k=1}^K 
\left(
\frac{1}{M N |B_k|}
\sum_{m=1}^M \sum_{i=1}^N 
\mathbf{1}_{B_k}(\bm{x}_i^{(m)}(t))
\right)
\mathbf{1}_{B_k}(\bm{x}),
\end{equation}
where $M$ is the number of realizations, $\bm{x}_i^{(m)}(t)$ is the position of the $i$th particle in the $m$th realization, and $\mathbf{1}_{A}$ is the indicator function of the set $A$.  
This construction \eqref{eq: rhod} yields a piecewise constant approximation of $\rho(\bm{x},t)$. One can also construct the macroscopic density using kernel-based estimators~\cite{silverman2018density}, which reduce sensitivity to bin alignment and typically produce smoother density profiles. As the number of particles $N$ and the number of independent realizations $M$ increase, and as the bin size or kernel bandwidth decreases in a compatible manner, these discrete approximations converge in a weak sense to the macroscopic density $\rho(\bm{x},t)$ associated with the mean field limit~\cite{sznitman2006topics,silverman2018density}.

It is nontrivial to solve the nonlocal aggregation diffusion equation \eqref{eq: pde_model} robustly, especially when the interaction forces are nonsmooth or when the dynamics develop singular densities.
For systems with gradient-flow structures, the authors in \cite{carrillo2019blob} introduce a blob method to regularize singular interaction kernels while remaining consistent with the underlying energy structure, and in \cite{carrillo2015finite}, the authors develop a structure-preserving finite-volume method that ensures positivity and energy dissipation (therein referred to as entropy).  
While these approaches provide robustness for nonsmooth interactions or singular densities, they often require small time steps or large particle ensembles to accurately restore macroscopic density profiles.

In this work, we use a pseudo-spectral method \cite{fornberg1998practical} to solve \eqref{eq: pde_model}, which extends the approach in \cite{wang2015noisyhegselmannkrausesystemsphase} to general interaction kernels.  
We approximate the density $\rho(\bm{x},t)$ by a truncated Fourier expansion,
\begin{equation}\label{eq:rhok}
\rho(\bm{x},t)
\approx
\sum_{\bm{k}\in\mathcal{K}}
\hat{\rho}_{\bm{k}}(t)\,
e^{\mathrm{i}\bm{k}\cdot \bm{x}},
\end{equation}
where $\mathcal{K}$ is a finite set of wave vectors determined by the spatial resolution.
Substituting \eqref{eq:rhok} into \eqref{eq: pde_model}, we obtain the governing equations for each Fourier coefficient
\begin{equation}\label{eq:rhok_ode}
\frac{d}{dt}\hat{\rho}_{\bm{k}}(t)
=
 \mathrm{i}\bm{k}\cdot \widehat{\rho(\bm{\theta}\star\rho)}_{\bm{k}}(t)
+ \mathrm{i}\bm{k}\cdot \widehat{\rho \bm{f}}_{\bm{k}}(t)
- \frac{\sigma^{2}}{2}\,|\bm{k}|^{2}\,\hat{\rho}_{\bm{k}}(t).
\end{equation}
We evaluate the nonlinear terms by first computing the products
$\rho(\bm{\theta}\star\rho)$ and $\rho\bm{f}$ in the physical space and then applying the Fourier transform.

We advance \eqref{eq:rhok_ode} in time using a semi-implicit scheme that treats the diffusion term implicitly and the nonlinear transport terms explicitly.  
Given the Fourier coefficients $\hat{\rho}_{\bm{k}}^{n}$ at time $t^{n}$, we compute the updated coefficients $\hat{\rho}_{\bm{k}}^{n+1}$ by solving the semi-implicit update
\begin{equation}\label{eq:semi_implicit}
\frac{\hat{\rho}_{\bm{k}}^{n+1}-\hat{\rho}_{\bm{k}}^{n}}{\Delta t}
=
 \mathrm{i}\bm{k}\cdot
\left(
\widehat{\rho^{n}(\bm{\theta}\star\rho^{n})}_{\bm{k}}
+
\widehat{\rho^{n} \bm{f}}_{\bm{k}}
\right)
- \frac{\sigma^{2}}{2}\,|\bm{k}|^{2}\,\hat{\rho}_{\bm{k}}^{n+1}.
\end{equation}
This semi-implicit scheme stabilizes the linear diffusion term through an implicit time advance, while preserving the efficiency of an explicit treatment for the nonlinear terms (i.e., the nonlocal and advective terms).

The pseudo-spectral method is particularly effective for high-dimensional systems and long-range interaction kernels. However, it applies directly only to periodic domains.
For problems with infinite domains or with no-flux boundary conditions, which often arise in kinetic equations, we instead consider a sufficiently large computational domain and impose periodic boundary conditions. We choose the domain size so that the density remains concentrated away from the boundary and decays to negligible values near the domain edges over the time interval of interest.
Using this strategy, we notice that the periodic limitation introduces minimal boundary artifacts from numerical experiments.

\subsection{Solving for the energy minimizers}
Recall that the sufficient conditions \eqref{eq: condition1a} and \eqref{eq: condition1b} yield the energy minimizer $\rho^*$. With the mass conservation, the local minimizer $\rho^*$ satisfies
\begin{equation}\label{eq: minimizer equation}
    \begin{aligned}
    \left(\rho^*\star\Theta\right)(\bm{x}) + F(\bm{x}) + \frac{\sigma^2}{2}\ln(\rho^*(\bm{x})) &= \lambda\\
    \int_{\Omega}\rho^*(\bm{x})\,d\bm{x} &= 1
    \end{aligned}
\end{equation}
with $\lambda \in\mathbb{R}$ being an unknown scalar. 
To ensure the positivity, we consider a transformation by letting $u = \ln(\rho^*)$. Therefore, \eqref{eq: minimizer equation} becomes
\begin{equation}\label{eq: transformed eqn}
   \begin{aligned}
    \left(e^u\star\Theta\right)(\bm{x}) + F(\bm{x}) + \frac{\sigma^2}{2}u(\bm{x}) &= \lambda,\\
    \int_{\Omega}e^{u(\bm{x})}\,d\bm{x} &= 1.
    \end{aligned}
\end{equation}
We discretize \eqref{eq: transformed eqn} on a set of collocation points $\{\bm{x}_i\}$ and replace the convolution term with numerical integration using the collocation points. This treatment converts \eqref{eq: transformed eqn} into a system of nonlinear equations characterized by unknown variables $\{u(\bm{x_i})\}$ and $\lambda$. We then solve for $\{u(\bm{x_i})\}$ and $\lambda$ using the Newton method.  
Once we have the numerical solutions of $\rho^*$, we further check if it satisfies the second-order constraint \eqref{eq: condition1b} to guarantee the stationary solution is a local minimizer. 

In the case that the interaction kernel $\Theta(\bm{x})$ corresponds to the Green's function of a linear differential operator $\mathcal{L}$, we can further convert the integral equation \eqref{eq: transformed eqn} into a differential equation, and solve the differential equation for $\rho^*$. Suppose that
\begin{equation}
    \mathcal{L}[\Theta](\bm{x}) = \delta(\bm{x}),
\end{equation}
where $\delta(\bm{x})$ is the Dirac delta function.
We apply $\mathcal{L}$ to both sides of the first equation in \eqref{eq: transformed eqn} and obtain
\begin{equation}\label{eq: differential-form of minimizer eqn}
    e^{u(\bm{x})} + \mathcal{L}[F](\bm{x}) + \frac{\sigma^2}{2}\mathcal{L}[u](\bm{x}) = \lambda \mathcal{L}[1].
\end{equation}
In Section~\ref{sec: laplace}, we use the Laplace potential $\Theta(x)=\exp(-|x|)$ as an example to further illustrate how to convert the nonlocal integral equation \eqref{eq: transformed eqn} into the differential equation \eqref{eq: differential-form of minimizer eqn}.

The energy minimizer $\rho^*$ is expected to coincide with a stable stationary density $\rho_{\infty}$ of the PDE model \eqref{eq: pde_model}.
We examine its spectral stability through a linear stability analysis of $\rho_{\infty}$.
We cast the right-hand side of \eqref{eq: pde_model} into the following form, upon discretization of the density $\rho$ using $N$ grid points in space,
\begin{equation}\label{eq: recast of minizer equation}
    \mathcal{F}\left(\vec{\rho}\right) = 0,
\end{equation}
where $\mathcal{F}$ denotes the vector field associated with the system \eqref{eq: pde_model} and $\vec{\rho} = \left(\rho_1,\rho_2,\ldots,\rho_N\right)^T$ with $\rho_j = \rho(x_j)$ for $j \in \left(1,2,\ldots,N\right)$.
Then, the stability analysis hinges on the spectrum of the corresponding Jacobian operator, which, upon discretization, becomes the matrix
\begin{equation}\label{eq: Jacobian matrix}
    J_F = \nabla \mathcal{F}\left(\vec{\rho}\right).
\end{equation}
One should thus compute the numerical eigenvalues of $J_F\left(\rho_\infty\right)$ where $\rho_\infty$ is the numerical steady state of \eqref{eq: pde_model}.
To avoid computing a full numerical Jacobian, one can utilize a numerically approximated one 
$J_F^a$, whose $j$th column, denoted as $\left(J_F^a\right)_{:,j}$, is given by
\begin{equation}\label{eq: numerical Jacobian}
    \left(J_F^a\right)_{:,j} = \frac{1}{2\varepsilon}\left(\mathcal{F}(\rho^* + \varepsilon e_j) - \mathcal{F}(\rho^* - \varepsilon e_j)\right),
\end{equation}
where $e_j$ is the basis vector whose $j$th slot is 1 and 0 elsewhere for $j \in \left(1,2.\ldots,N\right)$, and $0 < \varepsilon \ll 1$ is a smallness parameter. 
It is relevant to recall that only eigenvectors of this numerical Jacobian that integrate to $0$ over the domain of interest correspond to mass-conserving perturbations.

\section{Numerical experiments}
\label{sec: examples}
In this section, we present three representative examples to illustrate the contributions of this paper from different perspectives.
We first consider the noisy Hegselmann–Krause (HK) opinion model, where the interaction force is an indicator function. The discontinuous forces make steady states difficult to compute precisely, especially at low noise levels, since small deviations in opinions can lead to large jumps in the interaction force. We use one- and two-dimensional HK models to compare the microscopic model \eqref{eq: SDE_model} with the macroscopic model \eqref{eq: pde_model} and demonstrate their consistency.
Next, we study how stochastic noise affects density evolution in a model with competing attractive and repulsive forces. The noiseless version of this model has been studied in \cite{bernoff2011primer,leverentz2009asymptotic}, which identified three parameter regimes yielding distinct long-term behaviors.
Finally, we consider the Laplace interaction potential, which is the Green's function of a differential operator. In this case, the minimizer condition reduces from an integral equation to a differential equation. We compute the associated energy minimizers $\rho^*$ numerically and use this example to illustrate how the energy variational framework can be applied to study the long-term dynamics of the PDE model. We also examine the spectral stability of the long-term dynamics through linear stability analysis.

\subsection{The noisy HK model} 
We first consider the noisy HK model in opinion dynamics \cite{hegselmann2015opinion,pineda2013noisy}, in which 
\begin{equation}\label{eq: indicator kernel}
{\Theta}(\bm{x}) = \frac12|\bm{x}|^2\mathbf{1}_{|\bm{x}|<c}, 
\quad 
F(\bm{x})=0,
\end{equation}
where $|\bm{x}|=\|\bm{x}\|_2$ is $L_2$ norm in $\mathbb{R}^d$.  
In the HK model, $\bm{x}$ represents the opinions of agents.  
The model assumes that individuals compromise their opinions only if their opinion differences are less than a prescribed confidence bound $c$, while neglecting opinions outside this range.  
We consider both one- and two-dimensional versions of the HK model and compare density evolution from both the SDE model \eqref{eq: SDE_model} and the PDE model \eqref{eq: pde_model}. 

\begin{figure}[htp]
    \centering
    (a)  \hspace{4.55cm}  (b)  \hspace{4.55cm} (c) \\
    \includegraphics[width=0.99\linewidth]{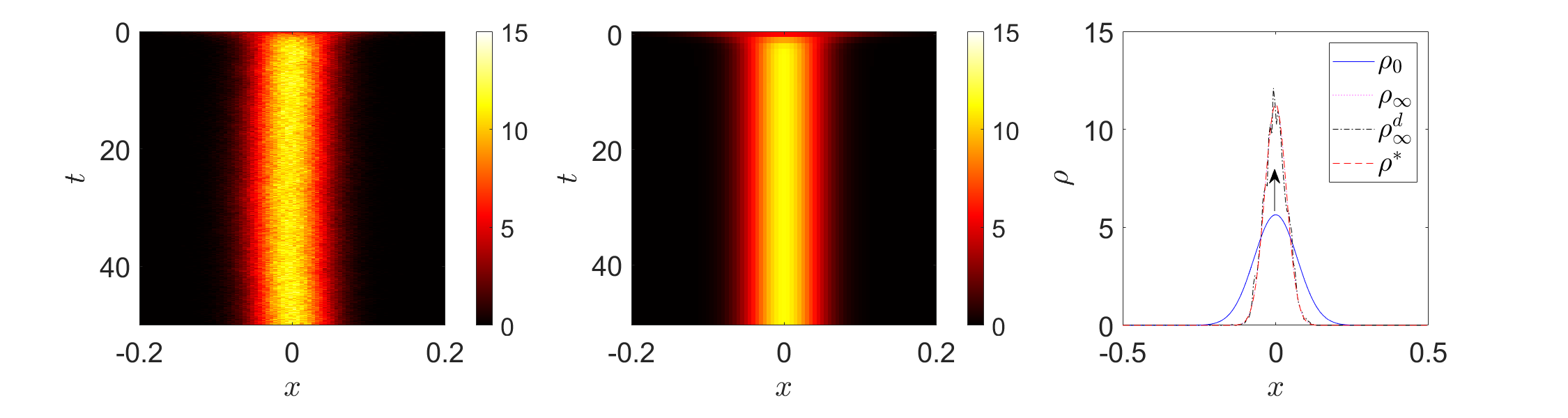}
    \caption{(a,b) The space-time density evolution for the 1D HK model for the discrete density $\rho^d$ in \eqref{eq: rhod} and the density $\rho$ from the PDE model \eqref{eq: pde_model}, respectively. (c) The initial density $\rho_0$, asymptotic densities $\rho_\infty,\rho_\infty^d$ (practically reached
    at $t=50$), and the energy minimizer $\rho^*$.  
    }
    \label{fig1: 1d case examples}
\end{figure}

Figure~\ref{fig1: 1d case examples} shows the numerical results for the one-dimensional HK model with parameters $c = 0.2$ and $\sigma = 0.05$, where panels (a) and (b) show the space-time evolution of the discrete density $\rho^d$ associated with \eqref{eq: SDE_model} and the density from the PDE model \eqref{eq: pde_model}, respectively. 
For the agent dynamics, we consider a system with $2000$ agents and sample the initial states independently from $\rho_0$, which is the initial condition of the PDE model \eqref{eq: pde_model}. To reduce the stochastic variations, we repeat the agent dynamics $2$ times and compute the average over all simulations.
The resulting density $\rho^d$ constructed via \eqref{eq: rhod} captures the overall trend of the continuum density from the PDE model but shows visible fluctuations due to randomness in the SDE model.
These fluctuations decrease as the number of agents increases and as more realizations are averaged to compute the ensembled empirical measure.
Figure~\ref{fig1: 1d case examples}(c) shows the density profiles at selected times, where $\rho_0$ is the initial density, $\rho_\infty$ and $\rho_\infty^d$ are the densities from the PDE and SDE model at $t=50$, for which we practically observe
that the system has reached a steady state, and $\rho^*$ is the energy minimizer obtained by solving \eqref{eq: minimizer equation}.
We notice that $\rho_\infty$ and $\rho^*$ are very close to each other, with $\rho^*$ almost overlapping with $\rho_\infty$, which agrees with the expectation that the gradient flow converges to the energy minimizer in the asymptotic time evolution limit.

We next consider the density evolution of the 2D HK model for the SDE and PDE models. 
In the 2D case, it is computationally expensive to sample a sufficiently large number of particles to ensure the discrete density $\rho^d$ provides a reliable approximation on every bin $\{B_k\}$ in \eqref{eq: rhod}.  
To address this issue, we apply a cubic spline interpolation using the data points $(\bm{\alpha}_k,\beta_k)$, where $\bm{\alpha}_k$ is the centroid of the bin $B_k$ and
\begin{equation}
\beta_k
=
\frac{1}{M N |B_k|}
\sum_{m=1}^M \sum_{i=1}^N 
\mathbf{1}_{B_k}\!\left(\bm{x}_i^{(m)}(t)\right)\neq 0.
\end{equation}
This interpolation smooths out sampling noise and provides a continuous approximation of the density suitable for comparison with the PDE solution.

\begin{figure}[htp]
    \centering
    (a)  \hspace{4.3cm}  (b)  \hspace{4.3cm} (c) \\
    \includegraphics[width=0.99\linewidth]{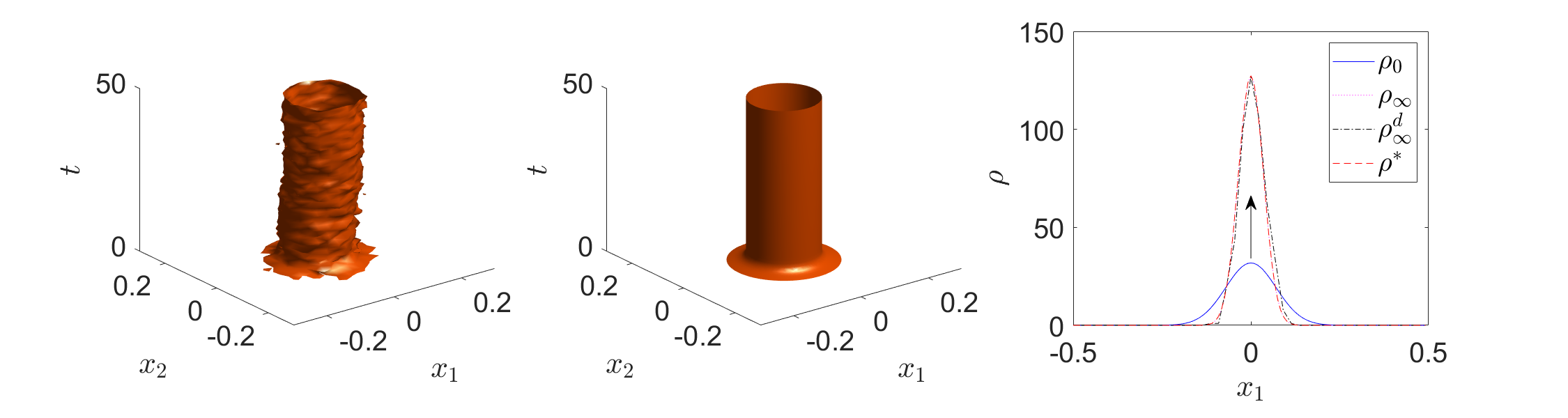}
    \caption{(a,b) Iso-surface plots for the discrete density $\rho^d$ from the SDE model and the density $\rho$ from the PDE 2D HK model at time $t=50$, where the asymptotic limit
    has been practically reached. 
    In the simulations, we use parameters $c = 0.25$ and $\sigma = 0.05$.
    (c) The cross section of the densities along the $x_2 = 0$ direction at $t=0$ and $t=50$ and the energy minimizer $\rho^*$ computed by solving \eqref{eq: minimizer equation}. 
    }
    \label{fig: 2d indicator kernel}
\end{figure}

Figure \ref{fig: 2d indicator kernel} shows the iso-surface plots of the discrete density $\rho^d$ constructed from the SDE model \eqref{eq: SDE_model} and the density $\rho$ from the PDE model \eqref{eq: pde_model}. 
The density value (either $\rho^d(x_1,x_2,t)$ or $\rho(x_1,x_2,t)$) remains constant for every triplet $(x_1,x_2,t)$ on the iso-surface plots, and the constant value is 5 in this example.
Figure \ref{fig: 2d indicator kernel}(c) shows the probability distributions $\rho(x_1,x_2=0,t)$ at $t=0$ and $t=50$, when we assume the dynamics have reached their steady states. 
In addition, we calculate the energy minimizer $\rho^*$ in the 2D case via a Newton method in two-dimensional space and plot the cross-section distribution $\rho^*(x_1,x_2=0)$ for reference. We observe that the energy minimizer $\rho^*$ overlaps with the steady-state density $\rho_\infty$ for the PDE model \eqref{eq: pde_model}, once again confirming our theoretical 
expectation that the gradient flow converges to a local energy minimizer. 
We expect similar agreement between the discrete density $\rho^d$ constructed from the SDE model \eqref{eq: SDE_model} and the density $\rho$ from the PDE model for other interaction potentials, including the additional examples considered in this section. We omit the corresponding numerical simulations for brevity.

\subsection{The Morse-type model with competing forces}
We next consider a Morse-type interaction potential with competing attractive and repulsive forces,
\begin{equation}\label{eq: Morse kernel}
\Theta(x) = -G \mathrm{e}^{-|x|/L} + \mathrm{e}^{-|x|},
\quad
F(x)=0,
\end{equation}
where $G>0$ and $L>0$ are the strength and characteristic length of attraction, and the second exponential term represents repulsion~\cite{mogilner2003mutual,leverentz2009asymptotic}.  
Leverentz et al.~\cite{leverentz2009asymptotic} classify the parameter pair $(G,L)$ into three regimes for the noiseless model (i.e., $\sigma=0$) based on the long-term behavior of solutions: a spreading regime (long-range repulsion dominant), in which densities disperse to infinity; a steady-state regime, in which densities converge to a localized aggregation; and a blow-up regime, in which densities concentrate into Dirac delta functions.
We choose parameters $G,L$ from each regime and study the corresponding model with noise level $\sigma=0.2$. In particular, we use $G=0.5$ and $L=0.5$ for the spreading case, $G=0.4$ and $L=4$ for the steady-state case, and $G=2$ and $L=0.5$ for the blow-up case.

\begin{figure}
    \begin{center}
    (a)  \hspace{4.35cm}  (b)  \hspace{4.35cm} (c) \\
     \includegraphics[width=0.8\linewidth]{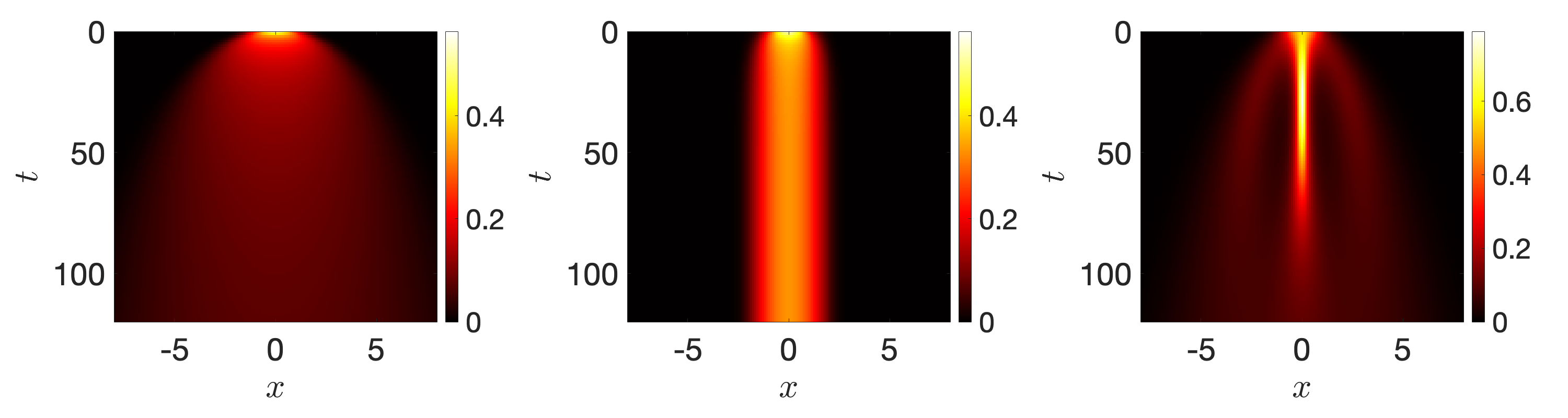} \\
        (d)  \hspace{4.3cm}  (e)  \hspace{4.3cm} (f) \\
     \includegraphics[width=0.8\linewidth]{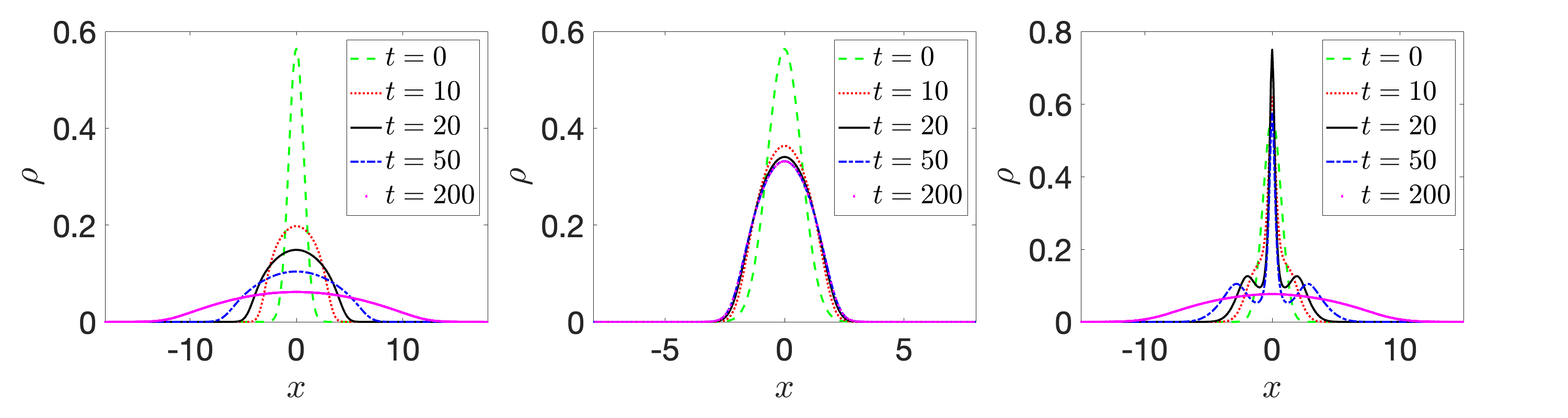}
    \end{center}
    \caption{Density evolution for the model with Morse potentials~\eqref{eq: Morse kernel} for (a,d) the spreading case, (b,e) the steady-state case, and (c,f) the blow-up case, where the top panels show the space-time evolution while the bottom panels show the densities at specific times. 
    }
    \label{fig: morse}
\end{figure}

Figure \ref{fig: morse} shows the space-time evolution of the densities together with density profiles at selected times.
For the spreading and steady-state cases, the densities either disperse into infinity or stabilize into localized aggregation, showing behavior qualitatively consistent with the noiseless model. 
In contrast, in the blow-up case, the density first forms several localized peaks (up to $t=70$) and then gradually smooths and merges into a single broader profile that continues to spread. This behavior differs from the noiseless setting, where the density converges to a finite sum of Dirac delta functions. 

\subsection{Laplace potential}
\label{sec: laplace}
In this subsection, we consider a 1D example with the potentials
\begin{equation}
\Theta(x) = e^{-|x|}, \quad F(x) = \gamma |x|^2,
\end{equation}
where $\gamma$ is a constant. Notice that the interaction potential $\Theta$ is a Laplace distribution and is the Green's function associated with the differential operator $\mathcal{L} = \partial_{x}^2 - 1$, satisfying 
\begin{equation}
\mathcal{L}\Theta(x) = -2\delta(x).
\end{equation}
We apply the differential operator $\mathcal{L}$ to the both sides of the first equation in \eqref{eq: condition1a} and obtain
\begin{equation}\label{eq: differential eqn by Green's function}
    \begin{aligned}
    -2\rho^*(x) + \partial_x^2F(x) - F({x}) + \frac{\sigma^2}{2}\left(\partial_x^2 - 1\right)\left(\ln\left(\rho^*(x)\right)\right) &= -\lambda,
    \end{aligned}
\end{equation}
which converts the integral equation in \eqref{eq: condition1a} into a differential equation. 

\begin{figure}[htp]
    \centering
    (a)  \hspace{4.55cm}  (b)  \hspace{4.55cm} (c) \\
    \includegraphics[width=0.8\linewidth]{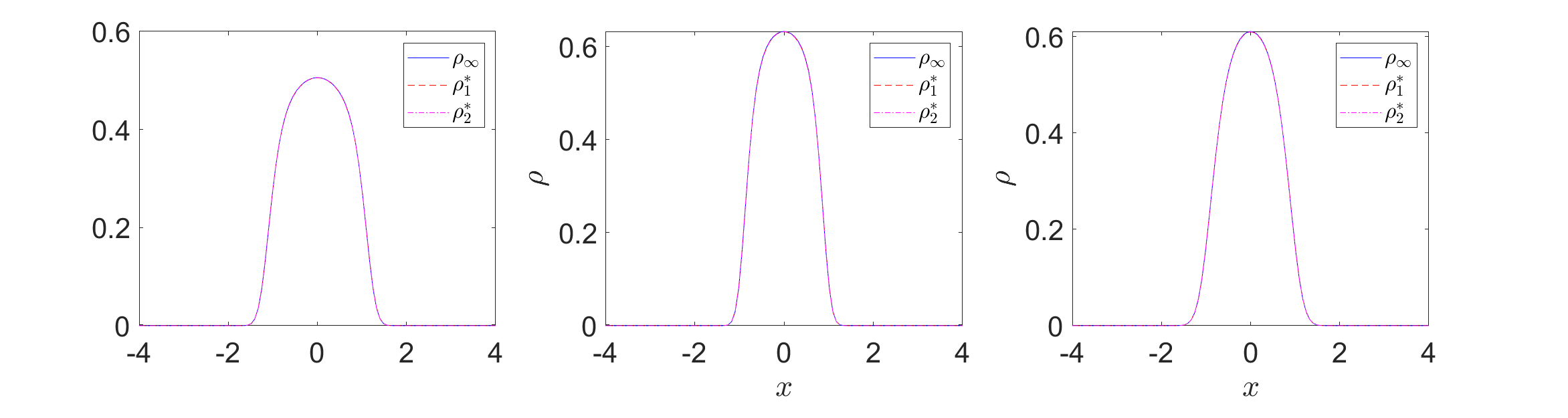}
    \caption{Long-term densities of \eqref{eq: pde_model} and energy minmizers solved from \eqref{eq: condition1a} and \eqref{eq: differential eqn by Green's function} for parameters (a) $\sigma=0.2$, $\gamma = 0.2$, (b) $\sigma = 0.2$, $\gamma = 0.3$, and (c) $\sigma =0.3$, $\gamma = 0.3$.
    }
    \label{fig: laplace}
\end{figure}

Figure \ref{fig: laplace} shows the steady-state solution $\rho_\infty$ from the PDE model \eqref{eq: pde_model},
based on asymptotic time evolution dynamics, together with the energy minimizers $\rho_1^*$ and $\rho_2^*$ obtained by solving the integral equation \eqref{eq: condition1a} and the differential equation \eqref{eq: differential eqn by Green's function}, respectively.
We consider several choices of the parameters $\sigma$ and $\gamma$. 
For all parameter settings, the global minimizers computed from either the integral formulation or the differential formulation (with the two of them being mathematically equivalent) coincide with the steady-state solutions obtained by long-term simulation of the PDE model \eqref{eq: pde_model}.
We also observe parameter-dependent trends.
As $\gamma$ increases, the density concentrates more strongly near the center, reflecting the increasing influence of the quadratic self-potential.
As $\sigma$ increases, diffusion becomes stronger and the density profile becomes smoother and more rounded.

In all cases, we also test the spectral stability of the corresponding density configurations by computing the eigenvalues of the Jacobian operator, which in practice is the discrete Jacobian matrix. 
We have examined various standard benchmarks thereof, such as the presence of a zero eigenvalue when the ODE pertaining to the density was translationally invariant (e.g., when $F(x)=0$). 
In all examples, the spectrum of the equilibrium density lies in the left half plane, confirming spectral stability and the minimizer nature of the configurations considered, which have also been found to be attractors of the corresponding energy-minimizing dynamics.
Since this serves only as an additional check of the energy-minimizing nature of the configurations, we do not present detailed associated results herein for brevity.

\section{Summary \& Outlook}
\label{sec: summary}
In this work, we investigate noisy nonlocal aggregation models for interacting particle systems, where both pairwise interactions and external forces arise as gradients of potentials. While deterministic nonlocal aggregation models have been extensively studied, the combined effects of nonlocal interactions and stochastic noise on long-term behavior, equilibrium structure, and stability are less well understood, especially from a unified microscopic–macroscopic and variational perspective.

We formulated the model at both microscopic and macroscopic levels. We began with a stochastic interacting particle system described by SDEs and then derived its continuum limit in the large population limit, which is an integro-differential equation.
Under a gradient structure assumption on the interaction and external forces, the system has an associated energy functional that incorporates nonlocal interactions, external potentials, and an entropy term induced by noise. 
We showed that the macroscopic model has a gradient flow formulation in the Wasserstein-2 space and satisfies an energy dissipation law. 
We further provided an energy variational framework to determine equilibrium states and analyze long-term dynamics. We derived first and second order variational conditions for energy minimizers in the probability space by considering admissible perturbations that satisfy positivity and mass conservation.

On the numerical side, we introduced methods to solve both the microscopic and macroscopic models and proposed procedures to reconstruct population densities from stochastic particle simulations. Using one- and two-dimensional noisy HK opinion models, we demonstrated the consistency between the microscopic and macroscopic density descriptions.
In addition, we considered the Morse-type interaction kernels with competing forces and studied how the stochastic noise influences the dynamics evolution of the macroscopic model in different regimes of parameters. 
We observed that the diffusion term helps stabilize the density evolution by competing with the attractive forces. In the blow-up parameter regime, the noise smooths the steady-state densities, which would otherwise concentrate into Dirac delta functions in the absence of noise.
We also considered a model with the Laplace potential, for which the variational conditions reduce to differential equations rather than the integral equations that typically arise in nonlocal aggregation models. 
We computed the long-term dynamics and the associated energy minimizers and found good agreement between them. In addition, we performed linear stability analysis, which confirmed the spectral stability and minimizer nature of these configurations and showed that they act as attractors of the corresponding energy minimizing dynamics.

Several directions for future work remain open. In this paper, we considered only a limited set of interaction kernels, while many others merit investigation, particularly with respect to their scaling properties and dimension-dependent behavior. 
Importantly, identifying broad classes of kernels for which general conclusions can be reached may be of particular interest.
Extending the analysis to noisy versions of these kernels would provide further insight into how stochastic effects interact with nonlocal interactions across different spatial dimensions and how they may influence the appearance of phase and structural transitions. 
A more systematic and quantitative study of parameter dependence, especially the role of the noise level in stabilizing or localizing energy minimizers, also represents an important direction for future research. 
In this work, we consider additive noise in the agent-based model. An alternative modeling choice involves multiplicative noise, which can confine the dynamics to a finite interval and extend nonlocal aggregation models to settings with state-dependent fluctuations.
Finally, from a numerical perspective, the methods used here can be extended towards a broader toolbox for such problems. Extending the current numerical examples to broader parameter regimes and higher-dimensional settings would further clarify the robustness and applicability of the proposed framework.

\begin{acknowledgement}
WC was funded in part by the National Science Foundation through DMS-2514053.
This material is partially based upon work supported by the U.S. National Science Foundation under the award PHY-2408988 (PGK). This research was partly conducted while P.G.K. was 
visiting the Okinawa Institute of Science and
Technology (OIST) through the Theoretical Sciences Visiting Program (TSVP)
and the Sydney Mathematical Research Institute (SMRI) under an SMRI Visiting
Fellowship. 
This work was also supported by the Simons Foundation
[SFI-MPS-SFM-00011048, P.G.K.].
\end{acknowledgement}

\bibliographystyle{abbrv}
\bibliography{references}

\end{document}